\documentclass[epj]{svjour}

\usepackage{afterpage}
\usepackage{graphicx,subfigure}
\usepackage{amsmath,amssymb}

\begin{document}

\title{Sedimentation and polar order of active bottom-heavy particles}

\author{Katrin Wolff\inst{1} \and Aljoscha M. Hahn\inst{1} \and Holger Stark\inst{1}}

\institute{                    
 Institut f\"ur Theoretische Physik, Technische Universit\"at Berlin, 
Hardenbergstrasse 36, 10623 Berlin, Germany
}
\date{Received: date / Revised version: date}
%
\abstract{
Self-propelled particles in an external gravitational field have 
been shown to display both an increased sedimentation length and polar order 
even without particle interactions. Here, we investigate self-propelled 
particles which additionally are bottom-heavy, that is they feel a torque 
aligning them to swim against the gravitational field. For bottom-heavy 
particles the gravitational field has the two opposite effects of i) 
sedimentation and ii) upward alignment of the particles' swimming direction. We 
perform a multipole expansion of the one-particle distribution of 
non-interacting particles with respect to orientation and derive expressions for 
sedimentation length and mean particle orientation which we check against 
Brownian Dynamics simulations. For large strength of gravity or small particle 
speeds and aligning torque, we observe sedimentation with increased 
sedimentation length compared with passive colloids but also active colloids 
without bottom-heaviness. Increasing, for example, swimming speed the 
sedimentation profile is inverted and the particles swim towards the top wall 
of the enclosing box. We find maximal orientational order at intermediate 
swimming speeds for both cases of particles with bottom-heaviness and those 
without. Ordering unsurprisingly is increased for the bottom-heavy particles, 
but this difference disappears at higher levels of activity and for very high 
activities ordering goes to zero in both cases.
} 

\maketitle

\section{Introduction}

Actively moving particles, being intrinsically out of equilibrium, exhibit a 
number of novel phenomena including interesting collective behaviour. Local 
alignment leads to the formation of moving swarms 
\cite{vicsek_novel_1995,ginelli_large-scale_2010} or even turbulence 
\cite{wensink_meso-scale_2012} and hydrodynamic interactions may induce 
long-range order 
\cite{aditi_simha_hydrodynamic_2002,saintillan_instabilities_2008,baskaran_statistical_2009}. 
When particles slow down in the vicinity of others, this interplay between local 
particle density and speed leads to a motility induced phase separation
\cite{tailleur_statistical_2008,bialke_crystallization_2012,farrell_pattern_2012,fily_athermal_2012,cates_when_2012,redner_structure_2012}.
Autophoretic effects, where colloids move along (self-generated) gradients of 
fields such as concentration or temperature, give rise to collective behaviour 
reminiscent of chemotaxis in biological swimmers
\cite{golestanian_collective_2012,theurkauff_dynamic_2012}.

But even in dilute suspensions, where particle interactions can be neglected, a 
number of interesting properties have been observed 
\cite{tailleur_sedimentation_2009,palacci_sedimentation_2010,enculescu_active_2011,pototsky_active_2012}. 
Activity changes the diffusive behaviour 
\cite{berg_random_1993,howse_self-motile_2007,golestanian_anomalous_2009,romanczuk_active_2012}
and thus, for example, results in increased sedimentation lengths under gravitation 
\cite{tailleur_sedimentation_2009,palacci_sedimentation_2010,enculescu_active_2011}. 
While this effect can be mapped onto passive particles at higher temperatures,
activity also aligns propulsion directions of particles against a gravitational
field and orientational order can be found even without particle interactions
\cite{enculescu_active_2011,pototsky_active_2012}. This polar order is the 
combined effect of self-propulsion and a non-constant density profile and 
appears without a gravitational torque aligning particles.

Here, we investigate non-interacting particles, which in addition to their 
activity exhibit \emph{bottom-heaviness}, that is a particle's centre of mass is 
displaced from its geometric centre in such a way that the particle tends to 
``point upward'' and thereby swims against an external gravitational field. For 
strong gravity, bottom-heaviness dominates particle orientation but in the 
colloidal regime, where thermal diffusion is not negligible, we observe a 
combination of activity-induced and bottom-heavy orientational order. Starting 
from the Smoluchowski equation for particle position and orientation, we derive 
an effective sedimentation process where activity and bottom-heaviness both 
enter into an increased effective diffusion constant and a reduced effective 
gravitation strength. We compare the analytically derived position density 
profile and orientational distribution function with results obtained from 
direct simulations of the equivalent Langevin equations. Furthermore, we 
identify a region in parameter space where the sedimentation profile is inverted 
and particles gather at the top wall of the enclosing box and determine density 
profiles at small to moderate Peclet numbers. This is the regime relevant for 
colloidal systems used, for example, in 
\cite{palacci_sedimentation_2010,theurkauff_dynamic_2012}.

\section{Model system}

For bottom-heavy particles the centre of mass does not coincide with the 
geometric centre but is displaced toward the \emph{rear} part of the particle, 
where \emph{front} and \emph{rear} are defined with respect to the swimming 
direction of the particle. The displacement is measured by the distance $r_0$.
Gravity $\mathbf{g}$ acts on the centre of mass and the torque
$\boldsymbol\tau = -m r_0 \mathbf{p}\times\mathbf{g}$ tends to align the
swimming direction $\mathbf{p}$ along $-\mathbf{g}$.

The resulting overdamped Langevin equations for particle position $\mathbf{r}$
and particle orientation $\mathbf{p}$ ($|\mathbf{p}|=1$) therefore read
\begin{eqnarray}
 \dot{\mathbf{r}} & = &  \mathbf{v} = v_0 \mathbf{p} + \frac{m}{\gamma}\mathbf{g} + \sqrt{2D}\boldsymbol\xi
	      \label{eq:langev3Di}\\
 \dot{\mathbf{p}} & = & \mathbf{\omega} \times \mathbf{p} = 
	      \frac{m r_0}{\gamma_r}\left(\mathbf{g}\times\mathbf{p}\right)\times\mathbf{p}
	      + \sqrt{2 D_r}\boldsymbol{\zeta}\times\mathbf{p} 
	      \label{eq:langev3Dii}
\end{eqnarray}
in the Stratonovich interpretation. The deterministic drift terms $v_0 
\mathbf{p}$ and $m\mathbf{g}/\gamma$ give the swimming velocity of the active 
particle and the downward velocity due to gravity, respectively.  The 
deterministic angular drift velocity is caused by the gravitational torque. 
While $m\mathbf{g}/\gamma$ actually requires the buoyancy mass
$\Delta m=(\rho_\mathrm{colloid} - \rho_\mathrm{water})R^3$ (density mismatch 
with surrounding fluid), the parameter in $mr_0/\gamma_r$ is indeed mass $m$. 
Here we assume that either $\Delta m \approx m$ or redefine the parameter $r_0$ 
as $r_0 m/\Delta m$.

Both translational and rotational motion are subject to stochastic velocities 
$\sqrt{2D}\boldsymbol\xi$ and $\sqrt{2 D_r}\boldsymbol{\zeta}$ where 
$\boldsymbol\xi$ and $\boldsymbol{\zeta}$ each contain three independent 
Gaussian random variables with zero mean and unit variance. In this Letter we 
assume that all noise is of thermal origin. So, the translational ($D$) and 
rotational ($D_r$) diffusion coefficients are connected with the respective 
friction coefficients via the Einstein relations $D =k_B T / \gamma$ and 
$D_r =k_B T / \gamma_r$. For spherical particles rotational and translational 
friction are related as $\gamma_r=4/3\,\, R^2\gamma$ which, as the only source 
of randomness is thermal noise in the fluid, translates to $D_r = 3D / (4 R^2)$ 
for the diffusion coefficients. 

We define a coordinate system by setting $\mathbf{g}=-g\mathbf{e}_z$ and 
non-dimensionalise spatial and temporal variables by setting 
$\mathbf{r}'=\mathbf{r}/R$ and $t'=tD/R^2$ and omitting the prime. This 
introduces the following dimensionless numbers, which we shall use as control 
parameters: the active Peclet number or simply \emph{activity},
$\mathrm{Pe} = v_0 R/D$, the gravitational Peclet number 
$\alpha = mgR / (k_\mathrm{B}T)$, which is the particle radius over 
sedimentation length of the passive system, and excentricity or bottom-heaviness 
$r_0'=r_0/R$.

The equivalent Smoluchowski equation for the one-particle density 
$\rho(\mathbf{r}, \mathbf{p},t)$ depending on both position and orientation then 
is
\begin{eqnarray}
 \partial_t \rho & = & -\nabla\cdot\mathbf{J}_t - \mathcal{R}\cdot\mathbf{J}_r
 \nonumber \\ 
  & = & \nabla^2 \rho +\left(\alpha\mathbf{e}_z - 
  \mathrm{Pe}\mathbf{p}\right)\cdot\nabla \rho \nonumber \\ 
  & & + \frac{3}{4}\left( \mathcal{R}^2\rho + \alpha r_0\mathcal{R} 
  \cdot \left(\mathbf{e}_z\times\mathbf{p}\rho\right)\right)
	      \label{eq:smol}
\end{eqnarray}
with $\mathcal{R}=\mathbf{p}\times\nabla_\mathbf{p}$ and the translational and
rotational currents $\mathbf{J}_t=(\mathrm{Pe}\mathbf{p}-\alpha\mathbf{e}_z)\rho
-\nabla \rho$ and $\mathbf{J}_r = -3/4\alpha r_0(\mathbf{e}_z\times\mathbf{p})
\rho - 3/4\mathcal{R}\rho$. At the system boundaries the translational current 
has to vanish.

\section{Multipole expansion}

In order to derive an approximative solution $\rho(\mathbf{r},\mathbf{p},t)$ to
Eq.~(\ref{eq:smol}) in a systematic way, we use the multipole expansion
\begin{eqnarray}
 \rho(\mathbf{r},\mathbf{p},t) & = & \rho^{(0)}(\mathbf{r},t) + 
 \boldsymbol\rho^{(1)}(\mathbf{r},t)\cdot\mathbf{p} \nonumber \\
  & & + \boldsymbol\rho^{(2)}(\mathbf{r},t)\colon \left(\mathbf{p}\mathbf{p} - 
 \frac{1}{3}\mathbb{I}\right) + \ldots \label{eq:harmonicsexpan}
\end{eqnarray}
with coefficients $\rho^{(0)}(\mathbf{r},t)$, $\boldsymbol\rho^{(1)}(\mathbf{r},t)$, 
$\boldsymbol\rho^{(2)}(\mathbf{r},t)$ depending on position and time but not on
particle orientation. Since in our system of non-interacting sedimenting 
particles the only interesting dynamics takes place in $z$-direction, we assume 
that $\rho$ only depends on $z=r_3$ and $u=p_3=\cos\theta$. The multipole 
expansion for $\rho$ reduces to an expansion in Legendre polynomials,
\begin{eqnarray}
 \rho(z,u,t) & = & \rho^{(0)}(z,t) + \rho^{(1)}(z,t)u \nonumber \\
 & & + \rho^{(2)}(z,t) \left(u^2 - \frac{1}{3}\right) + \ldots , \label{eq:legendreexpan}
\end{eqnarray}
and the Smoluchowski equation~(\ref{eq:smol}) becomes
\begin{eqnarray} 
 \partial_t \rho & = & \partial_z^2 \rho +\left(\alpha - \mathrm{Pe}u\right)
  \partial_z \rho  \nonumber \\
  & & + \frac{3}{4}\left(\partial_u (1 - u^2) \partial_u \rho 
  + \alpha r_0\partial_u (1 - u^2)\rho\right)      \label{eq:smol2}.
\end{eqnarray}

The position-dependent density $\Omega(z,t)=\int_{-1}^1 \rho(z,u,) du$, which is 
an easy observable in experiments, is related to the zeroth-order expansion 
coefficient $\rho^{(0)}(z,t)$ simply as $\Omega(z,t)=2\rho^{(0)}(z,t)$. We begin 
with deriving an effective drift-diffusion process for density $\Omega(z,t)$ 
valid on time scales larger than the orientational decorrelation time 
$\propto D_r^{-1}$. It is of the form
\begin{equation}
 \partial_t \Omega = -\partial_z J_\Omega ,    \label{eq:Omega}
\end{equation}
where $J_\Omega = \alpha_\mathrm{eff} \Omega - D_\mathrm{eff} \partial_z \Omega$
is the effective density current with the drift and diffusive part.

We now derive equations for coefficients 
$\rho^{(0)}$, $\rho^{(1)}$ and $\rho^{(2)}$ from Eq.~(\ref{eq:smol2}) making use
of the orthogonality of Legendre polynomials.
We first integrate Eq.~(\ref{eq:smol2}) over $u$ to obtain
\begin{equation}
 \partial_t \rho^{(0)} = \partial_z^2\rho^{(0)} + \partial_z
  \left(\alpha\rho^{(0)} - \frac{\mathrm{Pe}}{3} \rho^{(1)}\right).\label{eq:rho0}
\end{equation} 
Multiplying Eq.~(\ref{eq:smol2}) by $u$ and then integrating over $u$ yields the 
equation for $\rho^{(1)}$
\begin{eqnarray}
 \partial_t  \rho^{(1)} & = & \partial_z^2\rho^{(1)} + \alpha\partial_z\rho^{(1)} 
 - \mathrm{Pe}\partial_z\left(\frac{4}{15} \rho^{(2)} + \rho^{(0)}\right)
 \nonumber \\
 & & -\frac{3}{2}\rho^{(1)} + \alpha r_0\left(\frac{3}{2}\rho^{(0)} - 
 \frac{1}{5}\rho^{(2)}\right)\label{eq:rho1}
\end{eqnarray}
The same procedure for $u^2-1/3$ gives the next order:
\begin{eqnarray}
  \partial_t \rho^{(2)} & = & \partial_z^2\rho^{(2)} + \alpha\partial_z\rho^{(2)} 
 - \mathrm{Pe}\partial_z\left(\frac{9}{35} \rho^{(3)} + \rho^{(1)}\right)
 \nonumber \\
 & & -\frac{9}{2}\rho^{(2)} + \alpha r_0\left(\frac{9}{4}\rho^{(1)} - 
 \frac{27}{70}\rho^{(3)}\right) .
\label{eq:rho2}
\end{eqnarray}
So, each equation couples $\rho^{(i)}$ to the next higher order $\rho^{(i+1)}$.
Eqs.~(\ref{eq:rho1}) and~(\ref{eq:rho2}) describe massive modes with 
$\partial_t\rho^{(i)}\propto\rho^{(i)}$, $i=1,2$, which relax faster than the 
slow mode $\partial_t \rho^{(0)}\propto\partial_z\rho^{(0)}$ from 
Eq.~(\ref{eq:rho0}). We therefore omit the time derivatives in 
Eqs.~(\ref{eq:rho1}) and~(\ref{eq:rho2}) while keeping the time dependence in
Eq.~(\ref{eq:rho0}) (cf. Refs.~\cite{golestanian_collective_2012,cates_when_2012}). 
Furthermore, we want to keep Eq.~(\ref{eq:Omega}) at the drift-diffusion level 
with at most second-order derivatives and therefore drop second-order 
derivatives in Eq.~(\ref{eq:rho1}) and second- and first-order derivatives in 
Eq.~(\ref{eq:rho2}). We are still left with an equation for $\rho^{(2)}$ that 
couples it to the next higher order $\rho^{(3)}$. This coupling is due to the 
bottom-heaviness of particles as would be the case with any rotational drift 
term. To close the relation, we assume $\rho^{(3)} \ll \rho^{(1)}$ and arrive at
\begin{equation}
 \rho^{(2)} = \frac{\alpha r_0}{2} \rho^{(1)}.\label{eq:rho2solve}
\end{equation}
We plug this result into Eq.~(\ref{eq:rho1}) and obtain an equation in 
$\rho^{(1)}$. Taking its derivative with respect to $z$ and neglecting
second-order derivatives gives 
$\partial_z\rho^{(1)} \propto \partial_z\rho^{(0)}$, which we re-insert into the 
equation for $\rho^{(1)}$ to obtain
\begin{equation}
 \rho^{(1)} = \frac{2}{3+(\alpha r_0)^2 / 5} \left((3A-\mathrm{Pe})
  \partial_z \rho^{(0)} + \frac{3}{2}\alpha r_0\rho^{(0)}\right)\label{eq:rho1solve}
\end{equation}
with $A=(\alpha^2 r_0 - 2(\alpha r_0)^2\mathrm{Pe} / 15)/(3+(\alpha r_0)^2 / 5)$.
Using this result in Eq.\ (\ref{eq:rho0}), finally gives 
\begin{equation}
 \partial_t \rho^{(0)} = D_\mathrm{eff} \partial_z^2 \rho^{(0)} - 
  \alpha_\mathrm{eff} \partial_z \rho^{(0)} \label{eq:rho0solve}
\end{equation}
with effective diffusion coefficient
\begin{equation}
 D_\mathrm{eff} = 1 + \frac{2\mathrm{Pe}}{3 + (\alpha r_0)^2 / 5}
\left(\frac{\mathrm{Pe}}{3} - A\right)
\label{eq:Deff}
\end{equation}
and effective gravity
\begin{equation}
 \alpha_\mathrm{eff} = \alpha \left(1 - \frac{r_0\mathrm{Pe}}{3 + 
 (\alpha r_0)^2 / 5}\right).
 \label{eq:alphaeff}
\end{equation}
We have thus identified the coefficients for the drift-diffusion equation  for 
$\Omega(z,t) = 2 \rho^{(0)}(z,t)$ in Eq.~(\ref{eq:Omega}). 

The steady state solution for $\Omega(z)$ is an exponential profile with inverse 
sedimentation length $\alpha_\mathrm{eff}/D_\mathrm{eff}$,
$\Omega \propto \exp(-\alpha_\mathrm{eff} z /D_\mathrm{eff})$. We are now in 
a position to judge when the approximation to neglect derivatives of 
$\rho^{(i)}$, which we made in the derivation, is justified. As 
$\rho^{(i)}\propto\exp(-\alpha_\mathrm{eff} z / D_\mathrm{eff})$ for $i=0,1,2$
(Eqs.~(\ref{eq:rho2solve}, \ref{eq:rho1solve}, \ref{eq:rho0solve})), 
this approximation is possible when the absolute value of the reduced inverse 
sedimentation length $|\alpha_\mathrm{eff}/D_\mathrm{eff}| < 1$. 
For $\alpha \approx 1$ passive colloids have a sedimentation length equal to the 
particle radius $R$ meaning that particles accumulate at the bottom wall. Then, 
$\mathrm{Pe} \propto v_0$ and thereby $D_{\mathrm{eff}}$ have to be sufficiently 
large, so that active particles can leave this densely packed colloidal phase.

Within the regime of our approximations, we see that $D_\mathrm{eff}$ increases 
quadratically with $\mathrm{Pe}$ in accordance with the well-known result for 
the diffusion of (non-bottom-heavy) active particles 
\cite{berg_random_1993,howse_self-motile_2007}. The aligning torque 
($\alpha r_0$) counteracts this trend as it tends to rectify particle swimming 
directions and thereby decrease the randomness in the active motion. For very 
large $\alpha r_0$ the diffusion coefficient reverts to the passive case of 
purely thermal noise as swimming becomes fully deterministic. The effective 
gravitational  strength $\alpha_\mathrm{eff}$ goes roughly linearly in $\alpha$. 
Both bottom-heaviness ($r_0$) and activity ($\mathrm{Pe}$) work against gravity 
by causing the particles to swim upward. For $\alpha_\mathrm{eff} < 0$, activity 
together with bottom-heaviness overcomes gravity and particles swim towards the 
top resulting in inverted density profiles. For $\alpha_\mathrm{eff}=0$ the two 
effects balance and the density profile becomes uniform or constant in $z$ (far
from the confining walls). The limits of our approximation again become visible 
for very large $r_0$ where Eq.~(\ref{eq:alphaeff}) predicts $\alpha_\mathrm{eff}$ 
to tend to $\alpha$ instead of $\alpha-\mathrm{Pe}$.

The swimming speed necessary to cancel sedimentation and to obtain a uniform 
density profile can also be calculated independently of the multipole expansion.
A solution of Eq.~(\ref{eq:smol2}) that is constant in $z$ corresponds to the 
Boltzmann distribution of the particle orientation under the aligning torque, 
which gives  $\rho(z,u)\propto\exp(\alpha r_0 u)$, also known as Mises-Fisher 
distribution. Since the bounding walls enforce that the average translational 
drift current along the $z$ direction is zero, one has 
$0=\mathrm{Pe} \langle u \rangle - \alpha$. Calculating $\langle u \rangle$ from 
the Boltzmann distribution, we obtain
\begin{equation}
 \mathrm{Pe}^* = \frac{\alpha}{\coth(\alpha r_0)-\frac{1}{\alpha r_0}}
\label{eq:pe*}
\end{equation}
for the activity $\mathrm{Pe}^*$ which cancels gravity so that a uniform density 
profile occurs. The approximate result for $\mathrm{Pe}^*$ found from 
Eq.~(\ref{eq:alphaeff}) by setting $\alpha_\mathrm{eff} =0 $ amounts to
\begin{equation}
 \mathrm{Pe}^* \approx \frac{3}{r_0} + \frac{\alpha^2 r_0}{5} .
\end{equation}
It corresponds to the first two terms of the Laurent series of 
Eq.~(\ref{eq:pe*}) in $\alpha r_0$.

\begin{figure}[ht]
  \centering
 \includegraphics[width=0.45\textwidth]{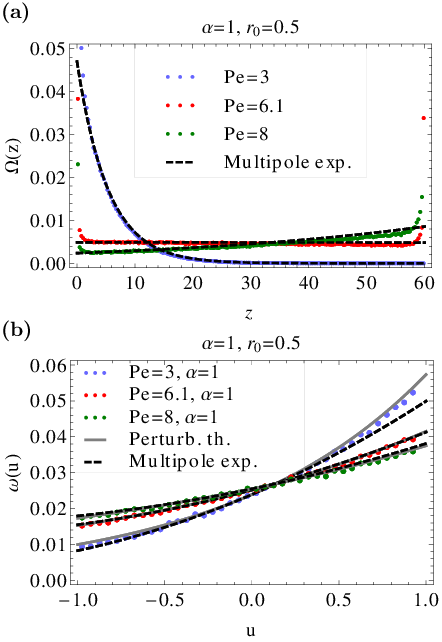}
  \caption{(Colour online) \textbf{(a)} Spatial density profile $\Omega(z)$ and 
    \textbf{(b)} orientational distribution $\omega(u)$ (within the bulk). 
    Numerical data (dots) are computed from Brownian Dynamics simulations of 
    Eqs.~(\ref{eq:langev3Di}) and (\ref{eq:langev3Dii}). Gravitational strength 
    $\alpha=1$ and bottom-heaviness $r_0=0.5$ are kept constant, activity is 
    $\mathrm{Pe}=3$ (blue dots), $\mathrm{Pe}=6.1\approx\mathrm{Pe}^*$ (red 
    dots) and $\mathrm{Pe}=8$ (green dots).
    }
  \label{fig:rho}
\end{figure}

\section{Sedimentation length and orientational order}

We solved the Langevin equations (\ref{eq:langev3Di}) and (\ref{eq:langev3Dii}) 
for the system of bottom-heavy active particles by performing Brownian Dynamics
simulations. We chose a cubic simulation box of size $L^3$ with $L = 60$ (in 
units of particle radii) with hard walls at $z=0$ and $z=L$ and periodic 
boundary conditions in $x$- and $y$-direction. For low levels of activity 
(see Fig.~\ref{fig:rho}~(a) for $\mathrm{Pe}=3$) the sole effect of 
bottom-heaviness is an increase in sedimentation length, while at higher 
activity, $\mathrm{Pe}=8$, bottom-heavy particles overcome gravity and produce 
inverted sedimentation profiles with particles swimming towards the top of the 
simulation box. The constant profile is realised for 
$\mathrm{Pe}=\mathrm{Pe}^*\approx6.1$. All three cases are captured very well 
by $\Omega(z)\propto\exp(-\alpha_\mathrm{eff}/D_\mathrm{eff})$ as can be seen 
in the comparison of simulations (dots) and theory (dashed lines) in 
Fig.~\ref{fig:rho}~(a). The only significant deviations between the two 
appear at the hard walls in $z$ which have been neglected in the theory. In the 
simulations, the boundary conditions in $z$ give rise to an accumulation of 
particles swimming against the wall and thus an increase in density next to the 
wall. The net orientation at the wall also differs from the bulk (not shown). 
This effect has been discussed in~\cite{enculescu_active_2011} and accumulation 
of colloids next to the walls has also been observed in 
experiments~\cite{palacci_sedimentation_2010}. 

The stationary orientational distribution function 
$\omega(u|z) = \rho(z,u) / \Omega(z)$ is found to be independent of position $z$ 
except in the vicinity of the bounding walls. We therefore write 
$\omega(u|z) = \omega(u)$ for short. The orientational distribution also depends 
on the particles' activity as can be seen from Fig.~\ref{fig:rho}~(b). This 
ordering is due to two effects: Firstly, the bottom-heaviness simply aligns 
particles against the vertical while the second effect due to activity works 
more subtly and in conjunction with the spatial density profile as shown in 
Ref.~\cite{enculescu_active_2011}. An excess of particles at position 
$z-\Delta z$ will mean that more particles reach position $z$ swimming upwards 
from below than swimming downwards from above and particles at $z$ will 
predominantly point upwards. At low activity, $\mathrm{Pe}=3$, the sedimentation 
profile is more pronounced and it produces alignment in the same direction as 
bottom-heaviness. In contrast, an inverted profile such as for $\mathrm{Pe}=8$ 
works in the opposite direction; it reduces orientational order and results in a 
flatter distribution, see Fig.~\ref{fig:rho}~(b). The orientational distribution 
$\omega(u)$ can also be derived from the multipole expansion 
Eq.~(\ref{eq:legendreexpan}). After insertion of $\rho^{(i)}$, $i=0,1,2$ we find
\begin{equation}
 \rho(u,z) = \Omega(z)\omega(u)
\label{eq:rhouz}
\end{equation}
 with
\begin{eqnarray}
 \omega(u) & \propto & 1 + \frac{1}{3+(\alpha r_0)^2 / 5}
 \left(\frac{\alpha_\mathrm{eff}}{D_\mathrm{eff}}(\mathrm{Pe}-3A) + \frac{3}{2}
  \alpha r_0\right) \nonumber \\
& & \times \left[2u + \alpha r_0 \left(u^2-\frac{1}{3}\right) \right]
\end{eqnarray}
As we only performed the expansion up to second order, $\omega(u)$ is quadratic 
in $u$ by construction. A better fit of the data, but a less systematic result, 
can be found from a perturbation ansatz in $\alpha$, 
$\omega(u) = \exp(\alpha f_1(u) + \alpha^2 f_2(u))$, where $f_1(u)$ and $f_2(u)$
can be determined from inserting Eq.~(\ref{eq:rhouz}) into Eq.~(\ref{eq:smol2}). 
Both results are plotted against simulation data in Fig.~\ref{fig:rho}~(b).

\begin{figure}[ht]
  \centering
    \includegraphics[width=0.4\textwidth]{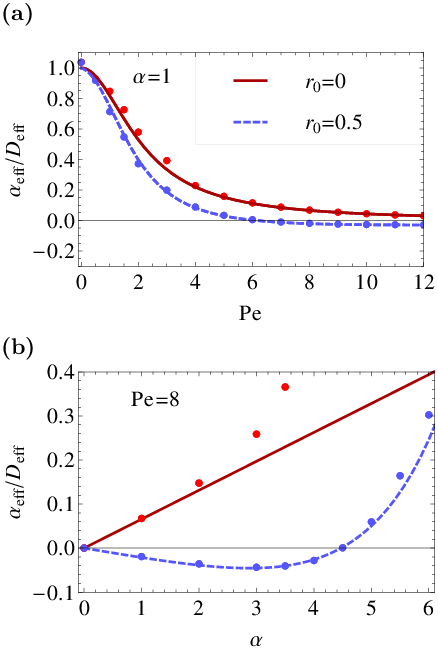}  
    \caption{(Colour online) Inverse sedimentation length 
    $\alpha_\mathrm{eff}/D_\mathrm{eff}$ \textbf{(a)} over activity 
    $\mathrm{Pe}$ at constant $\alpha=1$ and \textbf{(b)} over gravitational 
    strength $\alpha$ at constant $\mathrm{Pe}=8$. Lines are theoretical 
    predictions, dots are simulation results. Curves for non-bottom-heavy 
    particles ($r_0=0$) are shown in solid red, curves for bottom-heavy 
    particles ($r_0=0.5$) are dashed blue.
     }
  \label{fig:sedlength}
\end{figure}

\begin{figure}[ht]
  \centering
 \includegraphics[width=0.4\textwidth]{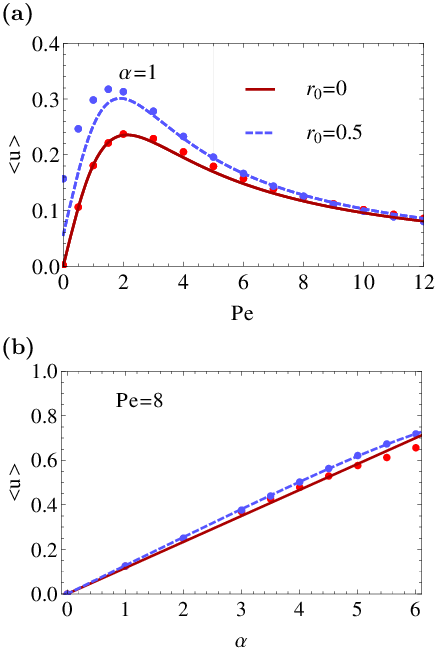}  
 \caption{(Colour online) Mean orientation $\langle u\rangle$ \textbf{(a)} over 
    activity $\mathrm{Pe}$ at constant $\alpha=1$ and \textbf{(b)} over 
    gravitational strength $\alpha$ at constant $\mathrm{Pe}=8$. Lines are 
    theoretical predictions, dots are simulation results. Curves for 
    non-bottom-heavy particles ($r_0=0$) are shown in solid red, curves for 
    bottom-heavy particles ($r_0=0.5$) are dashed blue.}
  \label{fig:meanu}
\end{figure}

We find that sedimentation lengths increase with activity $\mathrm{Pe}$
for both bottom-heavy and non-bottom-heavy particles. However, the increase for 
bottom-heavy particles is stronger as Fig.~\ref{fig:sedlength}~(a) demonstrates 
where we plot inverse sedimentation length $\alpha_\mathrm{eff}/D_\mathrm{eff}$ 
over $\mathrm{Pe}$. For large $\mathrm{Pe}$ both curves converge to 
$\alpha_\mathrm{eff}/D_\mathrm{eff}=0$ (infinite sedimentation length) as the
effective diffusion diverges quadratically in $\mathrm{Pe}$, that is the 
contribution of activity to random motion increases faster with $\mathrm{Pe}$ 
than the contribution to directed motion. However, the curve for 
bottom-heavy particles does so from below as for large $\mathrm{Pe}$ upswimming 
wins over sedimentation and the drift term becomes negative, 
$\alpha_\mathrm{eff}<0$. This regime of inverted profiles is visible even better 
in Fig.~\ref{fig:sedlength}~(b), where inverse sedimentation length is plotted 
against gravitational strength $\alpha$. For intermediate values of $\alpha$, 
bottom-heavy particles are oriented upwards but gravity is not yet strong enough
to pull particles down and we find maximally inverted profiles for 
$\alpha \approx 3$ (dashed blue curve). For non-bottom-heavy particles (solid 
red curve) the inverse sedimentation length increases linearly for small  
$\alpha$, in agreement with Ref.~\cite{enculescu_active_2011}. Deviations
between simulation and theory appear once sedimentation lengths become too small,
as discussed before. Parameter ranges relevant for active colloidal systems in 
Ref.~\cite{palacci_sedimentation_2010} are $\alpha=0.1$ and  $\mathrm{Pe}=0$ to 
$\mathrm{Pe}=6$ but larger values could be realised by increasing particle radii 
since $\alpha \propto R^4$ and $\mathrm{Pe} \propto R$. The region of inverted 
sedimentation for bottom-heavy particles is especially interesting as tuning of 
$\alpha$ in a centrifuge might be used to separate bottom-heavy from 
non-bottom-heavy colloids. 

As has been mentioned above, the orientational distribution changes with
activity $\mathrm{Pe}$ and becomes flatter for large values. Indeed, we
find a maximum in mean particle orientation 
$\langle u \rangle = \Omega(z)^{-1} \int_{-1}^1 \rho(u,z)u du$ for intermediate 
values of activity. From $\langle u \rangle = \rho^{(1)}/(3\rho^{(0)})$ we then 
find
\begin{equation}
 \langle u \rangle = \frac{2}{3+(\alpha r_0)^2/5}\left(
      \frac{\alpha_\mathrm{eff}}{D_\mathrm{eff}}
      \left(\frac{\mathrm{Pe}}{3} - A\right) +\frac{1}{2} \alpha r_0 \right)
\end{equation}
which at large $\mathrm{Pe}$ can be roughly viewed as an activity-induced part 
$\propto \alpha_\mathrm{eff}/D_\mathrm{eff}$ (although depending on the spatial 
profile which is changed by the presence of bottom-heaviness) and the direct 
contribution due to bottom-heaviness. In Fig.~\ref{fig:meanu}~(a) for fixed 
$\alpha=1$, polar order is maximal for $\mathrm{Pe}\approx2$ for 
non-bottom-heavy particles and for slightly lower values for bottom-heavy 
particles. Unsurprisingly, ordering is stronger for bottom-heavy particles as 
bottom-heaviness itself directly contributes to particle alignment. For very 
large activities, however, $\langle u \rangle$ tends to zero in both cases. For 
non-bottom-heavy particles this is simply explained by 
$\alpha_\mathrm{eff}/D_\mathrm{eff}$ going to zero as $\mathrm{Pe}^{-2}$. 
Physically, this means the spatial density profile $\Omega(z)$ becomes constant 
and activity-induced order 
$\propto \mathrm{Pe}\,\,\alpha_\mathrm{eff}/D_\mathrm{eff}$ vanishes. For 
bottom-heavy particles $\alpha_\mathrm{eff}/D_\mathrm{eff}$ goes to zero only 
as $\mathrm{Pe}^{-1}$ and activity-induced order becomes constant and cancels 
the bottom-heavy contribution.

Ordering increases linearly with $\alpha$ for small $\alpha$, see 
Fig.~\ref{fig:meanu}~(b), but flattens out for increasing values. For 
bottom-heavy particles the effect from increasing $\alpha$ is two-fold as it 
enhances both the passive alignment in an external potential and 
activity-induced alignment by sharpening the sedimentation profile. Mean 
orientation therefore increases with $\alpha$ faster for bottom-heavy 
particles than it does for non-bottom-heavy particles although at the relatively 
large activity of $\mathrm{Pe}=8$ the curves indicate that the main effect for 
orientational order comes from the sedimentation profile.

\section{Conclusion}
To conclude, we performed a multipole expansion up to second order in 
orientation for the one-particle density of active bottom-heavy particles
and determined a solution of the Smoluchowski equation in dilute suspensions 
which is prerequisite to the study of interacting particles. This approximation 
is useful in the colloidal regime where active motion and gravity are not too 
dominant compared with thermal diffusion and the regime explored here is 
experimentally accessible, e.g., by chemically powered Janus particles
\cite{palacci_sedimentation_2010,theurkauff_dynamic_2012}. Comparing 
orientational order in bottom-heavy and non-bottom-heavy particles, we 
elucidated the influence of both activity (cf. Ref.~\cite{enculescu_active_2011}) 
and of  bottom-heaviness.

We derived analytical expressions for the sedimentation length of bottom-heavy 
particles and discussed the two regimes of increased sedimentation lengths
\cite{tailleur_sedimentation_2009,palacci_sedimentation_2010,enculescu_active_2011} 
and of inverted sedimentation profiles. For fixed gravitational strength 
$\alpha$ or fixed activity $\mathrm{Pe}$, we find optimal values for the other 
parameter to maximise inversion. It is this density inversion and its 
instability under density fluctuations that, in the case of interacting 
particles, gives rise to pattern formation by 
bioconvection~\cite{childress_pattern_1975,pedley_new_1990,bees_linear_1998} 
(albeit in a parameter regime of large Peclet numbers where thermal diffusion is 
small compared with active motion). We are currently working on how this 
instability is induced by hydrodynamic interactions between individual swimmers.
Recently, it was demonstrated that in case of two Volvox algae hydrodynamic 
coupling leads to intricate bound states \cite{drescher_dancing_2009}.
Finally, we note that alignment of particles can also be achieved by mechanisms 
other than bottom-heaviness. For example, for superparamagnetic colloids 
\cite{ebert_experimental_2009} or magnetic Janus spheres \cite{yan_linking_2012} 
alignment is easily tuned by an external magnetic field whereas for biological 
swimmers chemotaxis or phototaxis result in similar 
phenomena~\cite{williams_tale_2011}.

\begin{acknowledgement}
We gratefully acknowledge financial support by the Deutsche 
Forschungsgemeinschaft through the Research Training Group GRK1558. 
\end{acknowledgement}

\bibliographystyle{epj}
\bibliography{BottomHeavinessWriteup}

\end{document}